# Electric field screening in atomically thin layers of $MoS_2$: the role of interlayer coupling

Andres Castellanos-Gomez[1,*], Emmanuele Cappelluti[2,3], Rafael Roldán[2], Nicolás Agraït[4,5], Francisco Guinea[2,*] and Gabino Rubio-Bollinger[4,*]

[1] Kavli Institute of Nanoscience, Delft University of Technology, Lorentzweg 1, 2628 CJ Delft (The Netherlands).
[2] Instituto de Ciencia de Materiales de Madrid. CSIC, Sor Juana Ines de la Cruz 3. 28049 Madrid, Spain.
[3] Institute for Complex Systems (ISC), CNR, U.O.S. Sapienza, v. dei Taurini 19, 00185 Rome, Italy
[4] Departamento de Física de la Materia Condensada. Universidad Autónoma de Madrid, Campus de Cantoblanco. E-28049 Madrid (Spain).
[5] Instituto Madrileño de Estudios Avanzados en Nanociencia IMDEA-Nanociencia. E-28049 Madrid (Spain).

E-mail: a.castellanosgomez@tudelft.nl, paco.guinea@icmm.csic.es, gabino.rubio@uam.es

Two dimensional crystals have recently emerged as an interesting family of materials with a large variety of electronic properties ranging from superconductors to topological insulators [1-8]. Although graphene is by far the most studied two-dimensional crystal [9], its lack of bandgap hampers its application in semiconducting and photonic devices. A large bandgap is a requirement, for instance, to fabricate field effect transistors with a large current on/off ratio and low power consumption. This fact has motivated the research in other 2D crystals with a large intrinsic bandgap such as atomically thin $MoS_2$ [10-16]. Single layer $MoS_2$ transistors have shown large in-plane mobility (200-500 $cm^2V^{-1}s^{-1}$) and high current on/off ratio (exceeding $10^8$) [17] making this material of great interest for electronic devices and sensors [17-19], possibly also in combination with graphene [20]. A deep insight into the charge distribution and on the electric field screening by atomically thin $MoS_2$ layers will allow to engineer $MoS_2$-based transistors towards an improved performance and also to understand the role of their layered structure in the electric field screening. However, no direct measurement of the electrostatic screening length in $MoS_2$ layers has been reported yet. Moreover, the role of the interlayer coupling in the screening (neglected for other layered materials such as few-layer graphene) is still unclear.

The aim of this work is to study the electrostatic screening by single and few-layer $MoS_2$ sheets by means of electrostatic force microscopy in combination with a non-linear Thomas-Fermi Theory to interpret the experimental results. We find that a continuum model of decoupled layers, which satisfactorily reproduces the electrostatic screening for graphene and graphite, cannot account for the experimental observations. A three-dimensional model with an interlayer hopping parameter can on the other hand successfully account for the observed electric field screening by $MoS_2$ nanolayers, pointing out the important role of the interlayer coupling in the screening of $MoS_2$.

**Figure 1**(a) shows an optical micrograph of a multilayered MoS$_2$ flake deposited on a Si/SiO$_2$ substrate. The regions showing different color under the optical microscope correspond to zones of the flake with different number of layers [21, 22], with the faint purple region at the center of Figure 1(a) corresponding to a single layer MoS$_2$. Figure 1(b) shows the AFM topography image, measured in contact mode in the region marked by a dashed square in Figure 1(a). A topographic line profile is also shown, indicating that the thickness of the thinner part of the flake is 0.7 nm which is compatible with the thickness of a MoS$_2$ single layer. Figure 1(c) shows the spatial map of the frequency difference between the E$^1_{2g}$ and A$_{1g}$ Raman modes, which increases monotonically with the number of MoS$_2$ layers [23, 24], measured in the area marked by a dashed rectangle in Figure 1(a). The frequency difference value clearly indicates that the number of MoS$_2$ layers is in good agreement with the values determined by AFM and the quantitative analysis of the optical contrast.

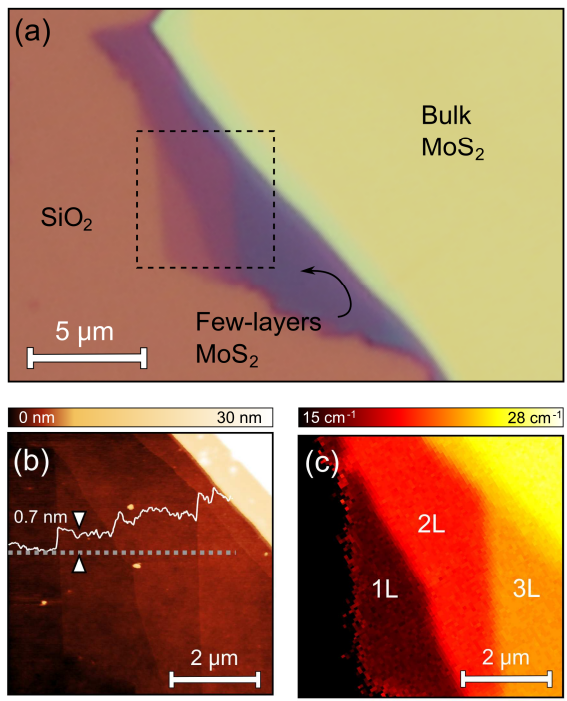

**Figure 1.** (a) Optical micrograph of a multilayered MoS$_2$ flake deposited onto a 285 nm SiO$_2$/Si substrate. (b) Topographic AFM image of the region marked by a dashed square in (a). A horizontal topographic line prolife is included in (b). (c) Spatial map of the frequency difference between the E$^1_{2g}$ and A$_{1g}$ Raman modes, also measured in the region marked by a dashed square in (a).

In order to gain a deeper insight into the interlayer screening in atomically thin MoS$_2$ flakes we employed electrostatic force microscopy (EFM) to probe the electric field, caused by charged impurities present in the MoS$_2$/substrate interface [25, 26], which is incompletely screened by the atomically thin MoS$_2$ crystals. The EFM measurements have been carried out as follows: the AFM tip is placed 20 nm above the surface of the flake and a voltage ramp is applied to the tip while measuring the oscillation amplitude of the cantilever which changes due to the tip-sample electrostatic force [26]. The relationship between the applied voltage and the electrostatic force is given by [27]

$$F = \frac{1}{2}\frac{\partial C}{\partial z}(V_{tip} - V_s)^2 \qquad (1)$$

where $C$ is the tip-sample capacitance at the tip-sample distance $z$, $V_{tip}$ is the tip-sample bias voltage and $V_s$ is the surface potential of the sample. The oscillation amplitude has a parabolic dependence on the tip-sample voltage and its vertex occurs at a voltage that counteracts $V_s$ (see **Figure 2**(a)). For bulk samples, the value of $V_s$ is just the tip-sample contact potential difference ($V_{CPD}$) due to the work function ($\Phi$) difference between tip and sample. For atomically thin MoS$_2$ samples, however, the electric field generated by charged impurities at the MoS$_2$/substrate interface cannot be fully screened and thus $V_s$ is modified (see sketch in Figure 2(b)). This effect can be seen as a shift of the parabola vertex as a function of the sample thickness, as shown in Figure 2(a). We have checked that the determined $V_s$ value does not depend on the tip-sample distance by repeating each measurement while increasing the tip-sample distance by small steps up to a total tip-sample distance of 40 nm. Additionally, we also observed that the determined value does not depend on the free-oscillation amplitude of the cantilever within the range employed in our experiment (5 nm to 2 nm).

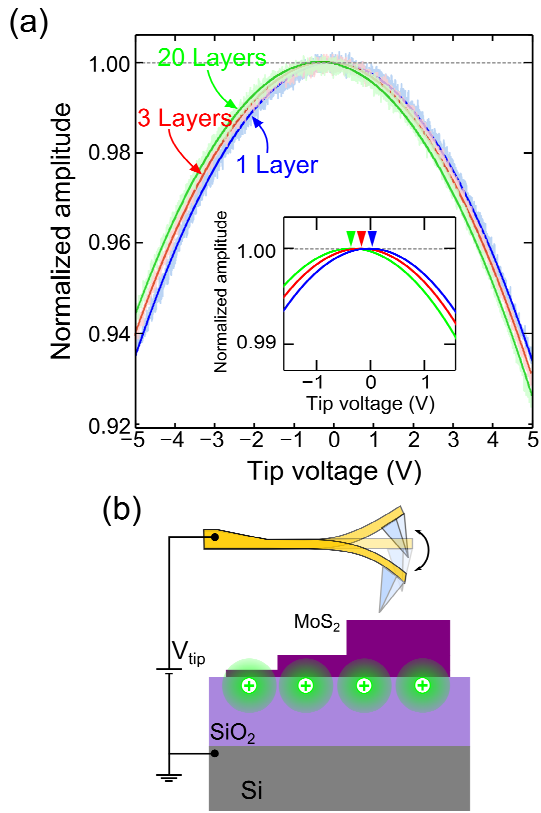

**Figure 2.** (a) Normalized cantilever oscillation amplitude as a function of the applied tip-sample bias voltage, measured in three regions with different number of layers. (inset) zoom in around the parabola maxima, only the fitting to the curves are shown to facilitate the identification of the apex (indicated by the triangles). (b) Schematic of the electrostatic force microscopy measurement setup.

**Figure 3** shows a systematic study of the dependence of the surface potential $V_s$ on the thickness of the MoS$_2$ nanosheets. While the presence of adsorbates on the surface may shift the surface potential, one can subtract this effect by considering the difference between the surface potential $V_s$ measured in a MoS$_2$ nanosheet and a thick MoS$_2$ flake (> 30 nm, considered as bulk). In this way one thus probes directly the out-of-plane electric field screening in atomically thin MoS2 crystals (See Figure 3(a) and 3(b)). For increasingly thick MoS$_2$ flakes the electric field generated by the charged impurities is increasingly screened and $V_s$ approaches the bulk value [28, 29]. For small thicknesses, the sign and magnitude of the deviation from the bulk value of $V_s$ ($\Delta V_s$) is related to the sign and the density of the charges in the MoS$_2$ flakes [29]. In particular, from the decrease of $\Delta V_s$ with flake thickness we infer negative (electron) doping, compatible with the presence of positively charged impurities in the substrate. The presence of positively charged impurities in the SiO$_2$ substrate is common in MoS$_2$ based field-effect transistor devices, showing a marked n-type behavior [17, 18, 30-33].

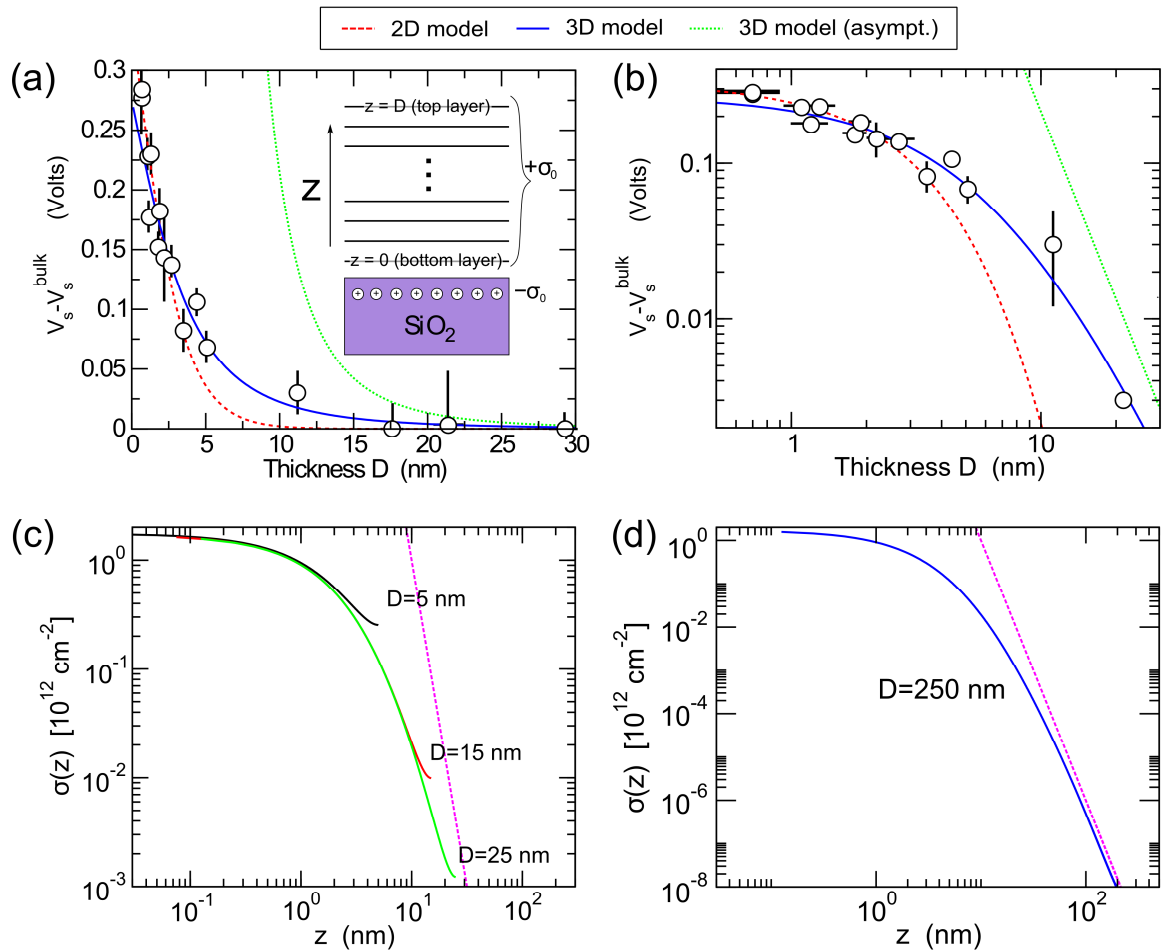

**Figure 3.** (a) Dependence of the deviation from the bulk value of the $V_s$ as a function of the flake thickness, caused by the electric field originated by charged impurities in the substrate which is incompletely screened by thin flakes. Lines are the theoretical predictions for a non-linear Thomas-Fermi 2D model of uncoupled MoS$_2$ layers (dashed red line), for the 3D model with z-axis dispersion (solid blue line), and for the asymptotic behaviour $\Delta V_s(D) \sim 1/D^4$ of this latter one. (inset) Sketch of the microscopic model. Vertical lines represent MoS$_2$ layers, while the variable $0<z<D$ runs over the thickness of the sample. (b) same as (a) with the axes in logarithmic scale to facilitate the comparison between the experimental data and the predictions with the different models. (c) Calculated surface charge distribution $\sigma(z)$ according with the 3D model across the samples for flakes with different thickness $D$. The dashed line represents the strong coupling asymptotic behavior $\sigma(z) \approx z^{-6}$. (d) shows how the strong-coupling regime is reached around 100 nm.

It is interesting to notice that the dependence of $\Delta V_s(D)$ as a function of thickness is quite weak, indicating a screening as poor as the one observed in graphene [29]. This appears to be quite surprising since the poor screening in graphene is related to the linear vanishing density of states (DOS) $N(\varepsilon) \sim \varepsilon$, whereas in MoS$_2$ the conduction band can be described by a conventional parabolic dispersion. Dimensionality plays thus here a non trivial role since an even weak hopping $t\perp$ between different layer of MoS$_2$ can change the low energy DOS from $N(\varepsilon) \sim$ const. to $N(\varepsilon) \sim \varepsilon^{1/2}$.

In order to gain a quantitative insight into this issue, we have employed a non-linear Thomas-Fermi (TF) theory for the screening properties, as successfully done for graphite [34, 35] and graphene [29]. Following a similar scheme as in Ref. [29], we first consider a continuum model of decoupled MoS$_2$ layers described by a parabolic two-dimensional (2D) conduction band $\varepsilon = \hbar^2 k_\parallel^2/2m_\parallel$. As experimental values for the in-plane mass $m_\parallel$ range in literature $m_\parallel \approx 0.01$-$0.1 m_e$ [36-38], the $m_\parallel$ value used here has been fitted (within this interval) to ensure the best agreement with the experimental data. We also take a dielectric constant along the $z$-axis $\kappa\perp = 7.4$ from Ref. [39]. We assume that a charge transfer takes place between the MoS$_2$ flakes and the SiO$_2$ substrate, leaving a net surface charge density $\sigma_0$ and establishing an underneath layer below the substrate surface with charge $\sigma_0$ (see inset in Figure 3(a)). The charge distribution $\sigma(z)$ as well as the electrostatic potential $V(z)$ in the multilayer samples as functions of $z$ (distance from the substrate) result thus from the energetic balance between the kinetic and the interlayer capacitance terms. Following Ref. [29], for a sample of thickness $D$, we introduce dimensionless parameter $r_D = \sigma(D)/\sigma(0)$. The screening properties are thus ruled by the implicit equation for $r_D$ for details see the Supplementary Material):

$$\int_{r_D}^1 \frac{du}{\left(u^2 - r_D^2\right)^{1/2}} = \sqrt{\frac{2\beta_0}{d}} D \Rightarrow \lim_{D \to \infty} r_D = 2\exp\left[-\sqrt{\frac{2\beta_0}{d}} D\right] \quad (2)$$

where $d = 6.14$ Å is the interlayer distance, and where the constant $\beta_0 = N_s N_v e^2 m_\parallel / 4\pi\varepsilon_0 \kappa\perp \hbar^2$ contains all the relevant parameters of the system, like the in-plane effective mass $m_\parallel$, the interlayer dielectric constant $\kappa\perp$, and the spin and valley degeneracies $N_s = N_v = 2$. Once determined $r_D$ from the implicit solution of Eq. (0.2), the potential drop $\Delta V(D)$ between the top and bottom layers is thus obtained as

$$\Delta V(D) = \frac{2\pi\hbar^2 \sigma_0}{N_s N_v m_\parallel} \sqrt{2\beta_0 d} \frac{1 - r_D}{\left(1 - r_D^2\right)^{1/2}} \quad (3)$$

One can easily see that $\Delta V(D)$ is expected to approach exponentially its asymptotic value $\Delta V(\infty) = \lim_{D \to \infty} \Delta V(D) = \left(2\pi\hbar^2 \sigma_0 / N_s N_v m_\parallel\right)\sqrt{2\beta_0 d}$.

The predictions of this model are also shown (red dashed line) in Figure 3(a), for $m_\parallel = 0.01 m_e$ and $\sigma_0 = 8 \times 10^{12}$ cm$^{-2}$. As we can see, this model reproduce in a satisfactory way the initial screening trend in samples with small thickness, but it is not able to account, due to its exponential behavior, the long tail for large $D$. We argue that the reduced screening properties of thick flakes of multilayer MoS$_2$ is essentially due to the interlayer hopping, which, unlike graphene, drives multilayer MoS$_2$ in a "weak-coupling" regime where the characteristic screening length $\xi$ is of the same order or larger than the sample thickness.

We can investigate this case by considering a three-dimensional (3D) model for bulk MoS$_2$ with an anisotropic three-dimensional conduction band $\varepsilon = \hbar^2 k_\parallel^2/2m_\parallel + \hbar^2 k_\perp^2/2m_\perp$, where the mass $m_\perp$ is related to the interlayer hopping $t_\perp$ as $m_\perp = \hbar^2/2|t_\perp|d^2$. We estimate $t_\perp \approx -0.2$ eV from the splitting of the minimum of the conduction band as a function of the number or layers,[40] and we get hence $m_\perp \approx 0.5 m_e$. Note also that in three-dimensional bulk MoS$_2$ the minimum of the conduction band shifts to a finite momentum along the K-Γ direction [41-43], so that the valley degeneracy results $N_v = 6$ [44]. Defining now $r_D = \sigma^{2/3}(D)/\sigma^{2/3}(0)$, the screening properties are now determined by the implicit equation

$$\left[\frac{25\beta_\perp d\sigma_0^2}{8(1-r_D^{5/2})}\right]^{-1/10} \int_{r_D}^1 \frac{du}{\left(u^{5/2}-r_D^{5/2}\right)^{1/2}} = \sqrt{\frac{2\beta_\perp}{d}}D \quad (4)$$

where here $\beta_\perp = (4e^2/5\varepsilon_0\kappa_\perp)(N_s N_v d m_\parallel \sqrt{m_\perp}/6\pi^2\hbar^3)^{2/3}$, and the difference potential across a sample of thickness $D$ (see Supplementary Material):

$$\Delta V(D) = \frac{1}{2}\left(\frac{6\pi^2\hbar^3}{N_s N_v d m_\parallel \sqrt{m_\perp}}\right)^{2/3}\left(\frac{25\beta_\perp d\sigma_0^2}{8}\right)^{2/5}\frac{1-r_D}{\left(1-r_D^{5/2}\right)^{2/5}} \quad (5)$$

The potential difference $\Delta V(D)$ for such three-dimensional model with the same parameters as before ($m_\parallel = 0.01 m_e$, $\kappa_\perp = 7.4$) and a slightly smaller charge density $\sigma_0 = 5\times 10^{12}$ cm$^{-2}$ are also shown in Figure 3(a) (solid blue line), with a remarkable improvement, in particular for what regards the long tail at large $D$. Note that the charge density $\sigma_0 = 5\times 10^{12}$ cm$^{-2}$ is very similar to what estimated for graphene samples on SiO$_2$,[21] suggesting that it is an intrinsic property (charged impurities) of the substrate. Note also that the reduced screening properties is not significantly related to the different analytical behavior in two and three dimensions of $\Delta V(D)$ in the asymptotic strong-coupling regime $D\to\infty$, although $\Delta V(D)$ changes from $\Delta V(D) \propto \exp(-D/\xi)$ in the 2D model to $\Delta V(D) \propto 1/D^4$ in the 3D model (dotted green line in Figure 3(a)). The main source of reduction of the screening properties is instead the the shift towards higher $D$ of the transition between weak and strong coupling regime. For the 2D model we would get such crossover for thickness $\overline{D} \approx 1$ nm, locating thus all the samples in the strong coupling regime, whereas the crossover is shifted for the 3D model to $\overline{D} \approx 30-50$ nm, so that essentially most of the samples investigated here are expected to be in the weak-coupling regime, with reduced screening properties. The change of regime between weak to strong-coupling is also better pointed out in Figure 3(c) where we report the charge density $\sigma(z)$ as function of the variable $z$ for samples with different thickness $D$. We can note that the strong-coupling power law behavior $\sigma(z) \approx z^{-6}$ is asymptotically recovered only for very large $z \geq 50$ nm and for very thick sample $D \geq 100$ nm (see Figure 3(d)). This is quite different from the graphene case where a strong-coupling regime is achieved already for $z \geq 1\text{-}2$ nm [29].

In conclusion, a combined experimental and theoretical study of the electrostatic screening by single and few-layer MoS$_2$ sheets has been presented. We have probed the electric field, generated by charged impurities in the MoS$_2$/substrate interface, which is incompletely screened by MoS$_2$ sheets with different number of layers. A three-dimensional non-linear Thomas-Fermi model with a non-negligible interlayer hopping parameter has been employed to reproduce the experimental results. This demonstrates that unlike for other atomically thin crystals such as graphene, the interlayer coupling plays an important role in the screening processes for MoS$_2$.

*Experimental*

MoS$_2$ nanosheets have been fabricated by mechanical exfoliation of MoS$_2$ (SPI Supplies, 429ML-AB) with Nitto Denko tape. In order to ensure the optical visibility of ultrathin MoS$_2$ layers, Si/SiO$_2$ wafers with a 285 nm SiO$_2$ layer are used [21]. We identify single and few layer MoS$_2$ sheets under an optical microscope and estimated the number of layers by their optical contrast [21]. Prior to the transfer, the Si/SiO$_2$ substrates have been cleaned following standard procedures in nanofabrication. First, the Si/SiO$_2$ substrate has been cleaned with nitric acid in an ultrasonic bath for 5 minutes. Second, the substrates have been thoroughly rinsed with deionized water. Third, the substrates have been immersed in iso-propanol and dried by blowing with nitrogen gas. In order to remove any organic residue from the surface, the substrates are further cleaned in a UV/Ozone generator (*Novascan*) for 10 minutes just before the transfer.

The topography of the MoS$_2$ nanolayers has been characterized with a *Nanotec Cervantes* AFM. Standard silicon cantilevers with spring constant of 40 N/m and tip curvature <10 nm have been used to operate in the amplitude modulation mode (for the EFM measurements). Softer cantilevers with spring constant of 0.7 N/m and tip curvature <20 nm have been employed to operate in the contact mode AFM. Contact mode AFM has been employed to avoid thickness determination artifacts due to the thickness dependent surface potential of the MoS$_2$ flakes. The AFM piezoelectric positioners have been calibrated by means of a recently developed method to provide accurate measurements of the flake thicknesses [45].

A micro-Raman spectrometer (*Renishaw in-via RM 2000*) was used in a backscattering configuration excited with a visible laser light ($\lambda = 514$ nm), at low power levels P < 1 mW, to double check the number of layers of the studied MoS$_2$ flakes [23, 43].

*Acknowledgements*

This work was supported by MICINN/MINECO (Spain) through the programs MAT2011-25046 and CONSOLIDER-INGENIO-2010 'Nanociencia Molecular' CSD-2007-00010, Comunidad de Madrid through program Nanobiomagnet S2009/MAT-1726 and the European Union (FP7) through the programs RODIN and ELFOS. E.C. acknowledges the Marie Curie Grant PIEF-GA-2009-251904.

This is the post-peer reviewed version of the following article:
A.Castellanos-Gomez *et al*. "Electric field screening in atomically thin layers of MoS$_2$: the role of interlayer coupling".
Advanced Materials, (2012). doi: 10.1002/adma.201203731
Which has been published in final form at:
http://onlinelibrary.wiley.com/doi/10.1002/adma.201203731/abstract
[1] K. S. Novoselov, A. K. Geim, S. V. Morozov, D. Jiang, Y. Zhang, S. V. Dubonos, I. V. Grigorieva, A. A. Firsov, Science 2004, 306, 666.
[2] C. Dean, A. Young, I. Meric, C. Lee, L. Wang, S. Sorgenfrei, K. Watanabe, T. Taniguchi, P. Kim, K. Shepard, Nature Nanotech. 2010, 5, 722.
[3] H. Steinberg, D. R. Gardner, Y. S. Lee, P. Jarillo-Herrero, Nano Lett. 2010, 10, 5032.
[4] A. Castellanos-Gomez, M. Wojtaszek, N. Tombros, N. Agraït, B. J. van Wees, G. Rubio-Bollinger, Small 2011, 7, 2491.
[5] A. Castellanos-Gomez, M. Poot, A. Amor-Amorós, G. Steele, H. van der Zant, N. Agraït, G. Rubio-Bollinger, Nano Res. 2012, 5, 550.
[6] D. Teweldebrhan, V. Goyal, A. A. Balandin, Nano Lett. 2010, 10, 1209.
[7] D. Teweldebrhan, V. Goyal, M. Rahman, A. A. Balandin, Appl. Phys. Lett. 2010, 96, 053107.
[8] J. Khan, C. Nolen, D. Teweldebrhan, D. Wickramaratne, R. Lake, A. Balandin, Appl. Phys. Lett. 2012, 100, 043109.
[9] K. S. Novoselov, D. Jiang, F. Schedin, T. J. Booth, V. V. Khotkevich, S. V. Morozov, A. K. Geim, Proceedings of the National Academy of Sciences of the United States of America 2005, 102, 10451.
[10] A. Ayari, E. Cobas, O. Ogundadegbe, M. S. Fuhrer, J. Appl. Phys. 2007, 101, 014507.
[11] A. Castellanos-Gomez, M. Poot, G. A. Steele, H. S. J. van der Zant, N. Agraït, G. Rubio-Bollinger, Adv. Mater. 2012, 24, 772.
[12] A. Castellanos-Gomez, M. Poot, G. Steele, H. van der Zant, N. Agrait, G. Rubio-Bollinger, Nanoscale Research Letters 2012, 7, 233.
[13] K. F. Mak, K. He, J. Shan, T. F. Heinz, Nature Nanotechnology 2012, 7, 494.
[14] T. Cao, G. Wang, W. Han, H. Ye, C. Zhu, J. Shi, Q. Niu, P. Tan, E. Wang, B. Liu, Nature Communications 2012, 3, 887.
[15] S. Kim, A. Konar, W. S. Hwang, J. H. Lee, J. Lee, J. Yang, C. Jung, H. Kim, J. B. Yoo, J. Y. Choi, Nature Communications 2012, 3, 1011.
[16] H. Wang, L. Yu, Y.-H. Lee, Y. Shi, A. Hsu, M. L. Chin, L.-J. Li, M. Dubey, J. Kong, T. Palacios, Nano Lett. 2012.
[17] B. Radisavljevic, A. Radenovic, J. Brivio, V. Giacometti, A. Kis, Nature Nanotechnology 2011, 6, 147.
[18] H. Li, Z. Yin, Q. He, H. Li, X. Huang, G. Lu, D. W. H. Fam, A. I. Y. Tok, Q. Zhang, H. Zhang, Small 2012, 8, 63.
[19] K. F. Mak, K. He, J. Shan, T. F. Heinz, Nat Nano 2012, advance online publication.
[20] L. Britnell, R. Gorbachev, R. Jalil, B. Belle, F. Schedin, A. Mishchenko, T. Georgiou, M. Katsnelson, L. Eaves, S. Morozov, Science 2012, 335, 947.
[21] A. Castellanos-Gomez, N. Agrait, G. Rubio-Bollinger, Appl. Phys. Lett. 2010, 96, 213116.
[22] H. Li, G. Lu, Z. Yin, Q. He, Q. Zhang, H. Zhang, Small 2012, 8, 682.
[23] C. Lee, H. Yan, L. E. Brus, T. F. Heinz, J. Hone, S. Ryu, ACS Nano 2010, 4, 2695.
[24] H. Li, Q. Zhang, C. C. R. Yap, B. K. Tay, T. H. T. Edwin, A. Olivier, D. Baillargeat, Adv. Funct. Mater. 2012.
[25] Y. Zhang, V. W. Brar, C. Girit, A. Zettl, M. F. Crommie, Nature Physics 2009, 5, 722.
[26] A. Castellanos-Gomez, R. H. M. Smit, N. Agraït, G. Rubio-Bollinger, Carbon 2012, 50, 932.
[27] T. Glatzel, S. Sadewasser, M. Lux-Steiner, Appl. Surf. Sci. 2003, 210, 84.
[28] N. J. Lee, J. W. Yoo, Y. J. Choi, C. J. Kang, D. Y. Jeon, D. C. Kim, S. Seo, H. J. Chung, Appl. Phys. Lett. 2009, 95, 222107.
[29] S. S. Datta, D. R. Strachan, E. J. Mele, A. T. C. Johnson, Nano Lett. 2009, 9, 7.
[30] S. Ghatak, A. N. Pal, A. Ghosh, ACS Nano 2011, 5, 7707.
[31] Z. Yin, H. Li, L. Jiang, Y. Shi, Y. Sun, G. Lu, Q. Zhang, X. Chen, H. Zhang, ACS Nano 2012, 6, 74.
[32] A. Castellanos-Gomez, M. Barkelid, A. M. Goossens, V. E. Calado, H. S. J. van der Zant, G. A. Steele, Nano Lett. 2012, 12, 3187.
[33] W. Choi, M. Y. Cho, A. Konar, J. H. Lee, G. B. Cha, S. C. Hong, S. Kim, J. Kim, D. Jena, J. Joo, Adv. Mater. 2012.
[34] S. Safran, D. Hamann, Phys. Rev. B 1981, 23, 565.

Supporting information:

# Electric field screening in atomically thin layers of MoS$_2$: the role of interlayer coupling

Andres Castellanos-Gomez [1,*], Emmanuele Cappelluti [2,3], Rafael Roldán [2], Nicolás Agraït [4,5], Francisco Guinea [2,*], Gabino Rubio-Bollinger [4,*]

[1] Kavli Institute of Nanoscience, Delft University of Technology, Lorentzweg 1, 2628 CJ Delft, The Netherlands.
[2] Instituto de Ciencia de Materiales de Madrid. CSIC, Sor Juana Ines de la Cruz 3. 28049 Madrid, Spain.
[3] Institute for Complex Systems (ISC), CNR, U.O.S. Sapienza, v. dei Taurini 19, 00185 Rome, Italy
[4] Departamento de Física de la Materia Condensada (C–III). Universidad Autónoma de Madrid, Campus de Cantoblanco, 28049 Madrid, Spain.
[5] Instituto Madrileño de Estudios Avanzados en Nanociencia IMDEA-Nanociencia, 28049 Madrid, Spain.

*E-mail: a.castellanosgomez@tudelft.nl , paco.guinea@icmm.csic.es , gabino.rubio@uam.es .

**Thomas-Fermi model**

A detailed description of the Thomas-Fermi model applied to evaluate the screening properties of multilayer graphene is provided in Ref. [1]. We outline here a suitable generalization for multilayer MoS$_2$ samples, both including and neglecting the interlayer hopping $t_\perp$. Apart for the interlayer hopping processes, another (slight) difference between the two cases is related to the different valley degeneracy. The latter case can be indeed described by parabolic conduction band with in-plane mass $m_\parallel$, $\varepsilon = \hbar^2 k_\parallel^2 / 2m_\parallel$, and located at the K point, hence with valley degeneracy $N_v = 2$ [2, 3]. On the other hand, the interlayer hopping not only yields an anisotropic mass, $\varepsilon = \hbar^2 k_\parallel^2 / 2m_\parallel + \hbar^2 k_\perp^2 / 2m_\perp$, but it also shifts the conduction minimum at an intermediate point along the K-Γ line, giving rise to a higher valley degeneracy $N_v = 6$ [3-6].

In both cases, we can write in a generic way the free energy in the continuum limit as:

$$\Omega = \int_0^D \frac{dz}{d} \left\{ E[\sigma(z)] - \mu\sigma(z) - \frac{ze^2\sigma_0}{2\varepsilon_0\kappa_\perp}\sigma(z) \right\} - \frac{e^2}{4\varepsilon_0\kappa_\perp} \int_0^D \frac{dz}{d} \int_0^D \frac{dz'}{d} \sigma(z)|z-z'|\sigma(z') \quad \text{(S.1)}$$

where $E[\sigma(z)]$ is the functional of the kinetic energy.

*1. Uncoupled MoS$_2$ layers ( $t_\perp = 0$ )*

In the uncoupled MoS$_2$ layers case, we get

$$E[\sigma(z)] = A_0 \sigma^2(z), \quad \text{(S.2)}$$

where

$$A_0 = \frac{\pi \hbar^2}{N_s N_v m_\parallel}, \qquad (S.3)$$

where, as we have mentioned, $N_v$ is the valley degeneracy (here $N_v = 2$), and $N_s = 2$ is the standard spin degeneracy. The screening properties are thus obtained by minimizing Eq. (1.1) with respect to $\sigma(z)$ and result to be ruled by the differential equation

$$\left[\frac{d\sigma(z)}{dz}\right]^2 - \frac{2\beta_0}{d}\sigma^2(z) = -\frac{2\beta_0}{d}\sigma^2(D), \qquad (S.4)$$

with the boundary conditions

$$\left.\frac{d\sigma(z)}{dz}\right|_{z=0} = 2\beta_0\sigma_0, \qquad \left.\frac{d\sigma(z)}{dz}\right|_{z=D} = 0. \qquad (S.5)$$

Here $\beta_0 = e^2/4\varepsilon_0\kappa_\perp A_0$ and $\sigma_0$ is the total areal charge density.

Eq. (1.4) can also be written in the convenient integral form

$$\int_{r_D}^{1} \frac{du}{\left(u^2 - r_D^2\right)^{1/2}} = \sqrt{\frac{2\beta_0}{d}} D, \qquad (S.6)$$

where $r_D = \sigma(D)/\sigma(0)$. The potential difference $\Delta V(D)$ between the two ends of a sample of thickness *D* results thus

$$\Delta V(D) = 2A_0\sigma_0\sqrt{2\beta_0 d}\,\frac{1-r_D}{\left(1-r_D^2\right)^{1/2}}. \qquad (S.7)$$

We can also write an analytical expression for the charge density $\sigma(z)$ as a function of the vertical coordinate for a given thickness:

$$\sigma(z) = -\sigma_0\sqrt{2\beta_0 d}\,\frac{\cosh\left[-\sqrt{\frac{2\beta_0}{d}}(z-D)\right]}{\sinh\left[\sqrt{\frac{2\beta_0}{d}}D\right]}. \qquad (S.8)$$

*1. Coupled MoS$_2$ layers ( $t_\perp \neq 0$ )*

One of the major effect of the interlayer hopping is to modify the functional form of the kinetic energy. In this case we can write

$$E[\sigma(z)] = A_\perp \sigma^{5/3}(z), \quad (S.9)$$

where

$$A_\perp = \frac{3}{10}\left(\frac{6\pi^2 \hbar^3}{N_s N_v d m_\parallel \sqrt{m_\perp}}\right)^{2/3}, \quad (S.10)$$

and where in this case $N_v = 6$. The differential equation ruling the screening properties reads thus in this case:

$$\left(\frac{df(z)}{dz}\right)^2 - \frac{2\beta_\perp}{d} f^{5/2}(z) = -\frac{2\beta_\perp}{d} f^{5/2}(D), \quad (S.11)$$

where $f(z) = \sigma^{2/3}(z)$ and where $\beta_\perp = 6e^2/25 A_\perp \varepsilon_0 \kappa_\perp$. In these notations, the boundary conditions read now:

$$\left.\frac{df(z)}{dz}\right|_{z=0} = \frac{5}{2}\beta_\perp \sigma_0, \qquad \left.\frac{df(z)}{dz}\right|_{z=D} = 0. \quad (S.12)$$

Defining in this case $r_D = f(D)/f(0)$, the implicit integral equation reads now:

$$\left[\frac{25\beta_\perp d\sigma_0^2}{8(1-r_D^{5/2})}\right]^{-1/10} \int_{r_D}^{1} \frac{du}{\left(u^{5/2} - r_D^{5/2}\right)^{1/2}} = \sqrt{\frac{2\beta_\perp}{d}} D. \quad (S.13)$$

We can also obtain the charge density $\sigma(z) = f^{3/2}(z)$ as a function of the vertical coordinate for a given thickness $D$ from the implicit solution:

$$\int_{f(0)}^{f(z)} \frac{df}{\left(f^{5/2} - f^{5/2}(D)\right)^{1/2}} = \sqrt{\frac{2\beta_\perp}{d}} z. \quad (S.14)$$

Eq. (0.14) admits an implicit solution as:

$$\frac{1}{f^{5/2}(D)}\left\{f(0)\sqrt{f^{5/2}(0) - f^{5/2}(D)}\, _2F_1\!\left[1,\frac{9}{10},\frac{7}{5},\frac{f^{5/2}(0)}{f^{5/2}(D)}\right]\right.$$
$$\left. -f(z)\sqrt{f^{5/2}(z) - f^{5/2}(D)}\, _2F_1\!\left[1,\frac{9}{10},\frac{7}{5},\frac{f^{5/2}(z)}{f^{5/2}(D)}\right]\right\} = \sqrt{\frac{2\beta_\perp}{d}} z, \quad (S.15)$$

where $_2F_1(x)$ is the Hypergeometric function. In the strong coupling regime $f(0) \gg f(z) \gg f(D)$ we can expand Eq. (0.15), to obtain $f(z) \propto z^{-4}$ and $\sigma(z) \propto z^{-6}$.

Finally, the potential difference $\Delta V(D)$ across a sample of thickness $D$ is obtained as:

$$\Delta V(D) = \frac{5A_\perp}{3}\left(\frac{25\beta_\perp d\sigma_0^2}{8}\right)^{2/5} \frac{1-r_D}{\left(1-r_D^{5/2}\right)^{2/5}}, \quad (S.16)$$

and, in the asymptotic limit $r_D \ll 1$, one obtains $\Delta V(D) \propto D^{-4}$.